\documentclass[preprint,review,12pt,authoryear]{elsarticle}

\usepackage{amssymb}

\usepackage{natbib}
\journal{JASTP}

\begin{document}

\begin{frontmatter}



\title{Reply to `Influence of cosmic ray variability on the monsoon rainfall and temperature': a false-positive in the field of solar---terrestrial research}


\author{Benjamin A. Laken}

\address{Postboks 1047 Blindern, 0316, Oslo, Norway}
\ead{blaken@geo.uio.no}
 
\begin{abstract}
A litany of research has been published claiming strong solar influences on the Earth's weather and climate. Much of this work includes documented errors and false-positives, yet is still frequently used to substantiate arguments of global warming denial. This manuscript reports on a recent study by Badruddin \& Aslam (2014), hereafter BA14, which claimed a highly significant ($p=1.4\times10^{-5}$) relationship between extremes in the intensity of the Indian monsoon and the cosmic ray flux. They further speculated that the relationship they observed may apply across the entire tropical and sub-tropical belt, and be of global importance. However, their statistical analysis---and consequently their conclusions---were wrong. Specifically, their error resulted from an assumption that their data's underlying distribution was Gaussian. But, as demonstrated in this work, their data closely follow an ergodic chaotic distribution biased towards extreme values. From a probability density function, calculated using a Monte Carlo sampling approach, I estimate the true significance of the BA14 samples to be $p=0.91$.

\end{abstract}

\begin{keyword}
Monsoon \sep Solar Variability \sep Cosmic ray flux \sep Statistics


\end{keyword}

\end{frontmatter}


\section{Introduction}\label{intro}
\citet{Bad14}, hereafter BA14, recently reported a solar---terrestrial link between the cosmic ray (CR) flux and the Indian Monsoon, which they suggested may have implications of global importance and support so-called `Cosmoclimatology' \citep{Svensmark07}. This work demonstrates the way in which their findings were erroneous.

\begin{figure}
	\centering
	\includegraphics*[scale=0.8]{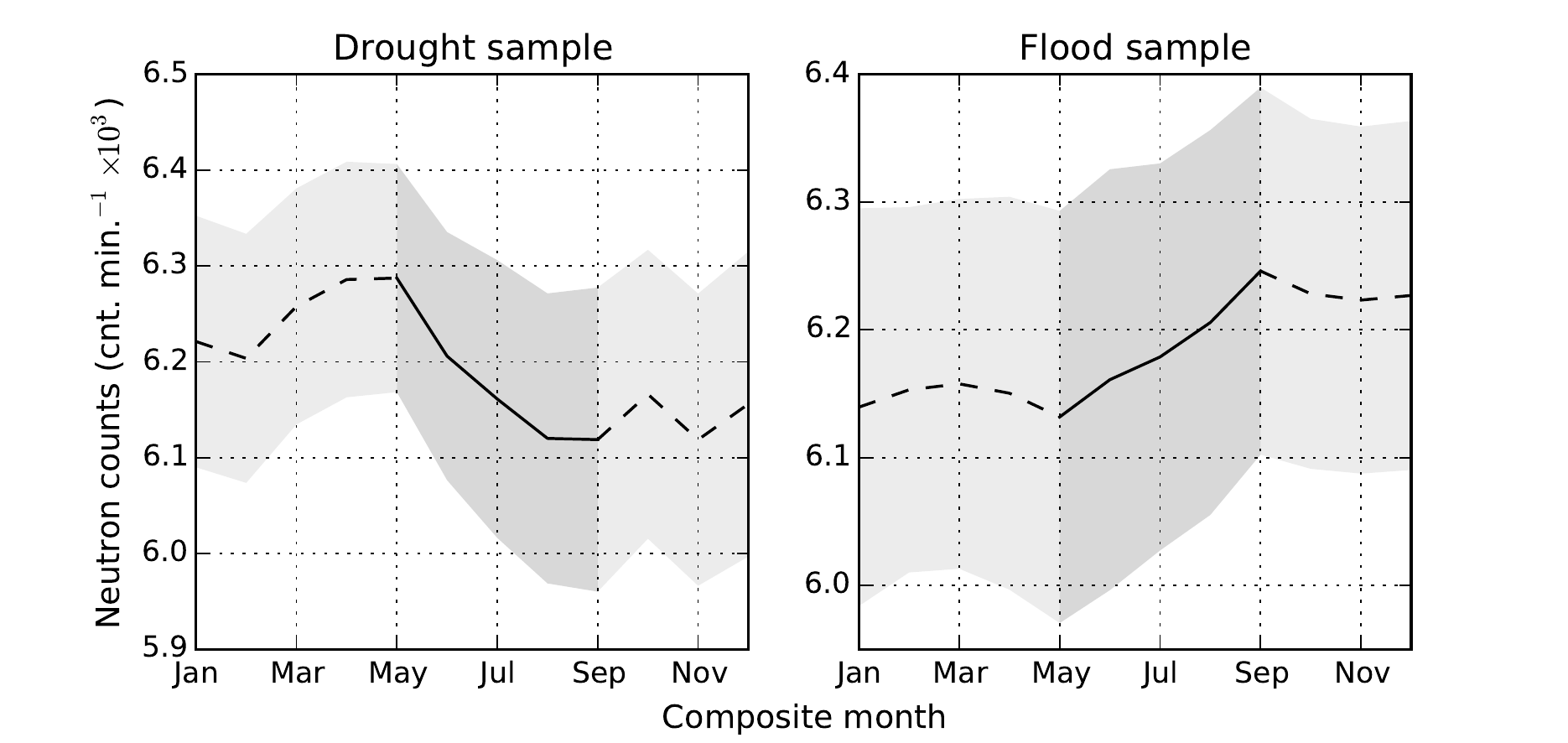} 
	\caption{Reproduction of the composite samples of BA14, showing the monthly-resolution pressure-adjusted neutron monitor count rate (units: counts min.$^{-1} \times 10^{3}$) from Oulu station (65.05$^{o}$ N, 25.47$^{o}$ E, 0.8 GV) occurring during 12 years of Indian monsoon `Drought' (D) and `Flood' (F) conditions. Composite means (in the matrix $m$-dimension) are plotted, with error ranges shown as $\pm$1 standard error of the mean (SEM) value. The period of May--September, selected by BA14 for Pearson's correlation analysis, has been emphasised in the plots.}
	\label{fig:composites}
\end{figure}

BA14 based their claims on highly significant statistical relationships obtained from composite (epoch-superposed) samples. Specifically, they examined linear changes in monthly neutron monitor counts, analysed over $m=5$ month periods (during the months of May--September) from two samples, each comprised of $n=12$ years of monthly resolution data: ie. composites from two matrices of $n \times m$ elements. The composites---which are vectors of the matrices averaged in the $n$-dimension---are respectively referred to as the `Drought' and `Flood' samples (which I shall also denote here as {\bfseries D} and {\bfseries F}), and represent the years of weakest and most intense monsoon precipitation respectively, recorded from 1964--2012. These data are shown in Figure ~\ref{fig:composites}, with the May--September periods of the composites emphasised: at first glance, it is true that these data show a linear change during the period highlighted, and also show anti-correlated between the {\bfseries D} and {\bfseries F} samples. 

Specifically, BA14 evaluated the Pearson's correlation coefficients ($r$-values) of the {\bfseries D} and {\bfseries F} samples, and used a standard two-tailed Student's t-test (which assumes a Gaussian distribution) to test their probability ($p$) values. They obtained values of $r=-0.95$ ($p=0.01$) for {\bfseries D}, and $r=0.99$ ($p=1\times10^{-3}$) for {\bfseries F}. Cumulatively, the $p$-value value was $1.4\times10^{-5}$: i.e. such a result should occur by chance only $1 / 71942$ times. BA14 interpreted these results to mean the lowest (and highest) precipitation volumes recording during the Indian monsoon period correspond to statistically significant decreases (and increases) in CR flux.

From these apparently highly significant CR flux changes, BA14 concluded that a solar---monsoon link exists, and operates via a theoretical CR flux cloud connection. They speculated that this connection impacts the monsoon in the following manner: increases in the CR flux enhance low cloud, rainfall, and surface evaporation, and also consequently decrease temperature (and vice versa). They further speculated that their findings may be expanded to the whole tropical and sub-tropical belt, and as a result may impact temperatures at a global scale. However, the significance of the {\bfseries D} and {\bfseries F} samples---and consequently the conclusions---of BA14 are wrong. This error resulted from the assumption of a Gaussian data distribution, which is not true of their data, as I shall demonstrate.

Moreover, of broader interest beyond the BA14 study is a recognition of a litany of fallacious solar---terrestrial studies: many of which have been re-examined in detail \cite[e.g. by][]{Pittock78, Pittock09, Farrar00, Krist00, Damon04, Sloan08, Calogovic10, Benestad09, Laken12, Laken13ERL}. False-positives within this field are of particular concern, as they contribute to a politically-motivated global warming denial movement. Providing material for groups intending to affect policy, such as the Heartland Institute's Nongovernmental International Panel on Climate Change (NIPCC) or the Centre for Study of Carbon Dioxide and Global Change \citep{Dunlap10}. Encouragingly though, a recent shift to open-access, and highly-repeatable workflows offers an opportunity for rapid communal development (and cross-checking) across a broad range of fields, including solar--terrestrial studies: at minimum, such approaches can more effectively facilitate the peer-review process, and enhance the quality and reliability of future publications. To illustrate this, this manuscript is supported by an accompanying iPython Notebook \citep{iPY}, enabling users of the open-source software to easily check, repeat, and alter the analysis. This notebook (and all accompanying data) are openly available from figshare \citep{Laken15}. 

\section{Analysis}\label{anal}

The CR flux oscillates between high and low values as solar activity progresses from minimum to maximum during the $\sim$11-year solar Schwabe cycle. Consequently, over the 5-month timescales with which the BA14 study was concerned, the CR flux spends relatively little time at stable values. As a result, the population of $r-$values which can be derived from 5-month composites of these data are ergodic and biased towards extreme values.

\begin{figure}
	\centering
	\includegraphics*[scale=0.7]{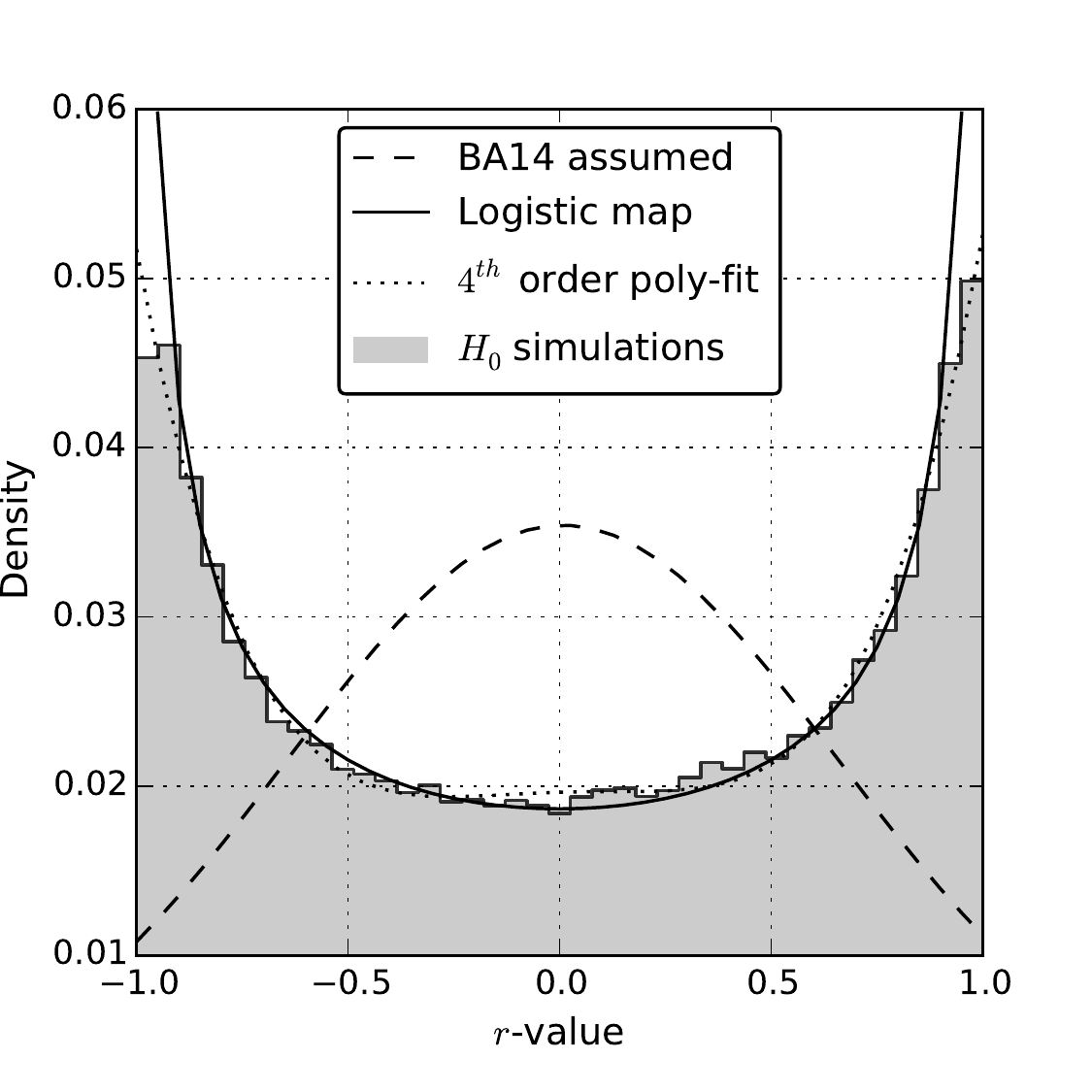} 
	\caption{Probability density function (PDF) of $r$-values drawn from 100,000 null hypothesis ($H_{0}$) composites (from Monte Carlo samples of the Oulu neutron monitor data). For comparison, a normalised Gaussian distribution (with a mean and standard deviation of $7.1\times10^{-1}$ and $6.5\times10^{-1}$) is plotted on the dashed line:  BA14 wrongly assumed the data possessed this distribution, and consequently, this was the source of their error. A logistical map, which predicts chaotic distributions (given in Equation ~\ref{eq:log}), is plotted on the solid line. A 4$^{th}$ order polynomial fit to the $H_{0}$ population is plotted on the dotted line: this fit can be used to calculate the probability ($p$) of a given $r$-value.}
	\label{fig:density}
\end{figure}

Using a Monte Carlo (MC) sampling approach, I have constructed 100,000 composites of equal dimensions to the original {\bfseries D} and {\bfseries F} samples from the neutron monitor data, and obtained $r$-values over May--September periods: I note that the May--September restriction is not strictly required, as in reality the only requirement is that the MC-samples span an identical time-period (5-months) to the original samples. I refer to these data as $H_{0}$ samples, as, by drawing these data randomly, they represent tests of the null hypothesis \citep[for more details on this method applied to solar---terrestrial studies see][]{Laken13}. A probability density function (PDF) of these data are presented in Figure ~\ref{fig:density}. For comparison, a normalised Gaussian distribution---assumed by BA14---is also shown (dashed line).

I have used two methods to model the PDF values: Firstly, a 4$^{th}$ order polynomial fit to the $H_{0}$ samples (shown on Figure  ~\ref{fig:density} as the dotted line), of the function $0.03625x^{4} - 0.0002797x^{3} - 0.0037x^{2} + 0.0007109x + 0.01964$. And secondly, analytically using a Logistic map \citep[as introduced by][]{May76}, which predicts the distribution of chaotically oscillating data (shown as the solid black line in Figure ~\ref{fig:density}). The formula for this is given in Equation ~\ref{eq:log} \citep{Ruelle89}, where $p$ is the probability, and $u$ is the variable (in this case $r$-values).

\begin{equation} \label{eq:log}
p(u)=\frac{1}{\pi\sqrt{1-u^{2}}}
\end{equation}

The distribution of $r$-values appear to follow the Logistic map to a high-degree, indicating that the solar-cycle is oscillating chaotically. Indeed, the chaotic nature of the solar cycle has been well described \citep[e.g.][]{Mundt91, Krem94, Rozelot95, Char01, Hans10, Hans13}. Disagreement between the Logistic map and the PDF occurs at the most extreme values, where $r$\textless$-0.9$ or $r$\textgreater$0.9$. 

\begin{equation} \label{eq:pval}
p=1-(0.03625x^{4} - 0.0002797x^{3} - 0.0037x^{2} + 0.0007109x + 0.01964)
\end{equation}

As the polynomial fit accurately follows the PDF, it can be readily used to estimate the $p$-value associated with a given $r$-value using Equation ~\ref{eq:pval}. From this, I calculate that the {\bfseries D} and {\bfseries F} samples possess $p$-values of 0.954 and 0.948 respectively, resulting in a cumulative $p$-value of 0.91, i.e. a chance of occurring under the null hypothesis of $1 / 1.1$. This result has a $p$-value four orders of magnitude larger than that estimated by the Student's t-test approach of BA14, and is virtually guaranteed by chance. Consequently, I conclude that the high $r$-values obtained in the BA14 composites do not support a relationship between extremes in Indian precipitation during the monsoon and co-temporal changes in the CR flux, but instead they are simply among the most commonly obtained values based on this sampling approach.

\section{Discussion}\label{diss}

The Cosmics Leaving OUtdoor Droplets (CLOUD) experiment at CERN has demonstrated that ion-mediated nucleation may lead to enhancements in aerosol formation of 2--10 times neutral values under specific laboratory conditions---low temperatures characteristic of the upper-troposphere, and with low concentrations of amines and organic molecules---however, this effect is absent under conditions more closely representing the lower troposphere \citep{Kirkby11,Almeida13}. Despite this, even if we assume that a significant nucleation of new aerosol particles form with solar activity, climate model experiments (which include aerosol microphysics schemes) have found that this would still not result in a significant change in either concentrations of cloud condensation nuclei or cloud properties. This is because the majority of the newly formed particles are effectively scavenged by pre-existing larger aerosols \citep{Pierce09, Snow11, Dunne12, Yu12}. These conclusions are supported by satellite and ground-based observations \citep[e.g.][]{Erlykin09, Kulmala10, Laken12Jclim, Benestad13, KT13}. For these and additional reasons, the IPCC AR5 concluded that the CR flux has played no significant role in recent global warming \citep{Boucher13}.

The numerous pitfalls into which solar---terrestrial studies in particular may fall, were lucidly outlined nearly 40-years ago by \citet{Pittock78}. Despite this, many studies with improper statistical methods, black-box approaches, and ad-hoc hypotheses still frequently appear. This problem is prominent within the field of solar---terrestrial studies. Consequently, the literature is replete with cases of demonstrated false-positives \citep[e.g.][]{Friis91, Marsh00, Shaviv03, Scafetta08, Svensmark09, Dragic11}, many of which have been (and continue to be) used as the basis for claims behind global warming denial \citep[e.g. such as in][]{Idso09, Idso13}, immediately making cases such as the one described in this manuscript a serious prospect in need of address.

\section{Acknowledgements}\label{ack}
I would like to thank Dr. Beatriz Gonz\'{a}lez-Merino (Instituto de Astrof\'{i}sica de Canarias), Dr. Ja\v{s}a \v{C}alogovi\'{c} (University of Zagreb), and Professors Frode Stordal and Joseph H. LaCase (University of Oslo), for helpful discussions. I also acknowledge the Python and iPython project (http://ipython.org). Data sources: Prof. Ilya Usoskin and the Sodankyla Geophysical Observatory for the Oulu neutron monitor data (http://cosmicrays.oulu.fi), and the Indian Institute of Tropical Meteorology (www.tropmet.res.in) for the monthly precipitation data (which, in this work, were taken directly from BA14).

\section{References}\label{refs}
\bibliography{B_refs}

\end{document}